\documentclass[%
  aps,prl,
  amsmath,amssymb,floatfix,%
  preprint,endfloats*
]{revtex4-1}

\usepackage[dvipdfm]{graphicx} 
\setkeys{Gin}{clip=true,draft=false}
\graphicspath{{./}{../eps/}}
\DeclareGraphicsExtensions{.eps,.ps}

\usepackage{dcolumn} 
\usepackage{bm} 

\usepackage{color}           
\usepackage[normalem]{ulem}  

\begin{document}

\title{Comment on ``Direct Measurement of Auger Electrons Emitted from a Semiconductor Light-Emitting Diode under Electrical Injection: Identification of the Dominant Mechanism for Efficiency Droop''
[Phys.\ Rev.\ Lett.\ 110, 177406 (2013)]%
}

\author{Francesco Bertazzi}
\affiliation{Dipartimento di Elettronica e Telecomunicazioni, Politecnico di Torino, corso Duca degli Abruzzi 24, 10129 Torino, Italy}
\affiliation{IEIIT-CNR, Politecnico di Torino, corso Duca degli Abruzzi 24, 10129 Torino, Italy}

\author{Michele Goano}
\email{michele.goano@polito.it}
\affiliation{Dipartimento di Elettronica e Telecomunicazioni, Politecnico di Torino, corso Duca degli Abruzzi 24, 10129 Torino, Italy}
\affiliation{IEIIT-CNR, Politecnico di Torino, corso Duca degli Abruzzi 24, 10129 Torino, Italy}

\author{Xiangyu Zhou}
\affiliation{Dipartimento di Elettronica e Telecomunicazioni, Politecnico di Torino, corso Duca degli Abruzzi 24, 10129 Torino, Italy}

\author{Marco Calciati}
\affiliation{Dipartimento di Elettronica e Telecomunicazioni, Politecnico di Torino, corso Duca degli Abruzzi 24, 10129 Torino, Italy}

\author{Giovanni Ghione}
\affiliation{Dipartimento di Elettronica e Telecomunicazioni, Politecnico di Torino, corso Duca degli Abruzzi 24, 10129 Torino, Italy}

\author{Masahiko Matsubara}
\affiliation{Department of Electrical and Computer Engineering, Boston University, 8 Saint Mary's Street, 02215 Boston, MA, USA}

\author{Enrico Bellotti}
\affiliation{Department of Electrical and Computer Engineering, Boston University, 8 Saint Mary's Street, 02215 Boston, MA, USA}

\date{\today}

\maketitle

In a recent Letter \cite{2013Iveland_PRL}, Iveland \emph{et al.} presented a careful spectroscopic study of the electrons emitted from the GaN $p$-cap (the highly Mg-doped layer between the active region and the anode contact) of a forward-biased InGaN/GaN light-emitting diode (LED).
Remarkably, the authors observed at least two distinct peaks in the electron energy distribution curves (EDCs), separated by about 1.5\,eV.

The lower-energy peak was attributed to photoemitted electrons excited in the band bending region (BBR, near the surface of the GaN $p$-cap) by the light generated in the active region.
(The work function $\Phi_\text{GaN}$ at the cesiated GaN surface was estimated at 2.3\,eV.
No information was provided in \cite{2013Iveland_PRL} about the peak wavelength of the electroluminescence spectrum but, according to \cite{2003Vurgaftman_JAP}, the energy gap of the eight In$_{0.18}$Ga$_{0.82}$N quantum wells (QWs) in the active region is about 2.5\,eV, and numerical solution of the Poisson--Schr{\"o}dinger equation \cite{2011Chiaria_JQE} suggests an effective gap of about 2.6\,eV between the fundamental levels and about 3.05\,eV between the second subbands of the conduction and valence bands (CB, VB).)

The authors concluded also that the only viable explanation for the higher-energy peak observed in the EDCs was the presence, at the $p$-cap surface, of an important population of high-energy electrons generated by Auger recombination processes in the active region and quickly thermalized at the bottom of an upper valley of the CB of GaN (labeled ``L valley'' in \cite{2013Iveland_PRL}).
The actual position in the Brillouin zone of this satellite valley and its energy above the bottom of the conduction band ($\Gamma_1^c$) was considered still somehow uncertain by the authors.
Critically, Iveland \emph{et al.} assumed that the electrons collected in this satellite valley would not undergo significant relaxation towards $\Gamma_1^c$ before reaching the BBR, so that the ``L valley'' would act as a source of high-energy electrons for emission into the vacuum.
According to \cite{2013Iveland_PRL}, no other mechanism could explain the observed EDCs, which therefore could be considered the first experimental demostration of the role of Auger recombination as an important loss mechanism responsible for LED efficiency droop.

Auger-induced leakage of high-energy carriers from the active region \cite{1988Chik_JAP.1} has been already suggested as a relevant effect in GaN-based LEDs \cite{2012Deppner_PSSR}, but Iveland \emph{et al.} go further, affirming that the signature of Auger processes may be \emph{directly} observed on the EDC after the carriers have drifted through the entire extension not only of the 40\,nm-thick Al$_{0.15}$Ga$_{0.85}$N electron blocking layer (EBL) delimiting the active region (with a conduction band offset $\Delta E_c \approx 200$\,meV with respect to the surrounding GaN layers, assuming $\Delta E_c / \Delta E_g \approx 0.67$), but also of the 200\,nm-thick GaN $p$-cap.
The description proposed in \cite{2013Iveland_PRL} hinges on three assumptions:
\begin{enumerate}
  \item the presence of a satellite valley in the CB at an energy from $\Gamma_1^c$ compatible with the observed higher-energy peak,
  \item a relaxation rate from that satellite valley to $\Gamma_1^c$ slow enough to allow an important fraction of electrons to reach the BBR before decaying to the bottom of the CB,
  \item negligible carrier heating by the electric field in the BBR, since, although electrons may gain kinetic energy in the BBR, their total energy at the surface would remain smaller than in the bulk.
\end{enumerate}
In the following we will discuss these three assumptions, and we will conclude that, according to our understanding of electronic structure and high-energy carrier transport in GaN, only the first one is probably correct.
(As a consequence, we will not consider another, stronger claim made in \cite{2013Iveland_PRL}, where a combined analysis of the measured EDCs and emitted light intensity is said to demonstrate unambiguously that Auger recombination is the dominant droop mechanism.)

\section{Local minima in the GaN conduction band}

\begin{figure*}
\centerline{\includegraphics*[width = 0.70\linewidth]{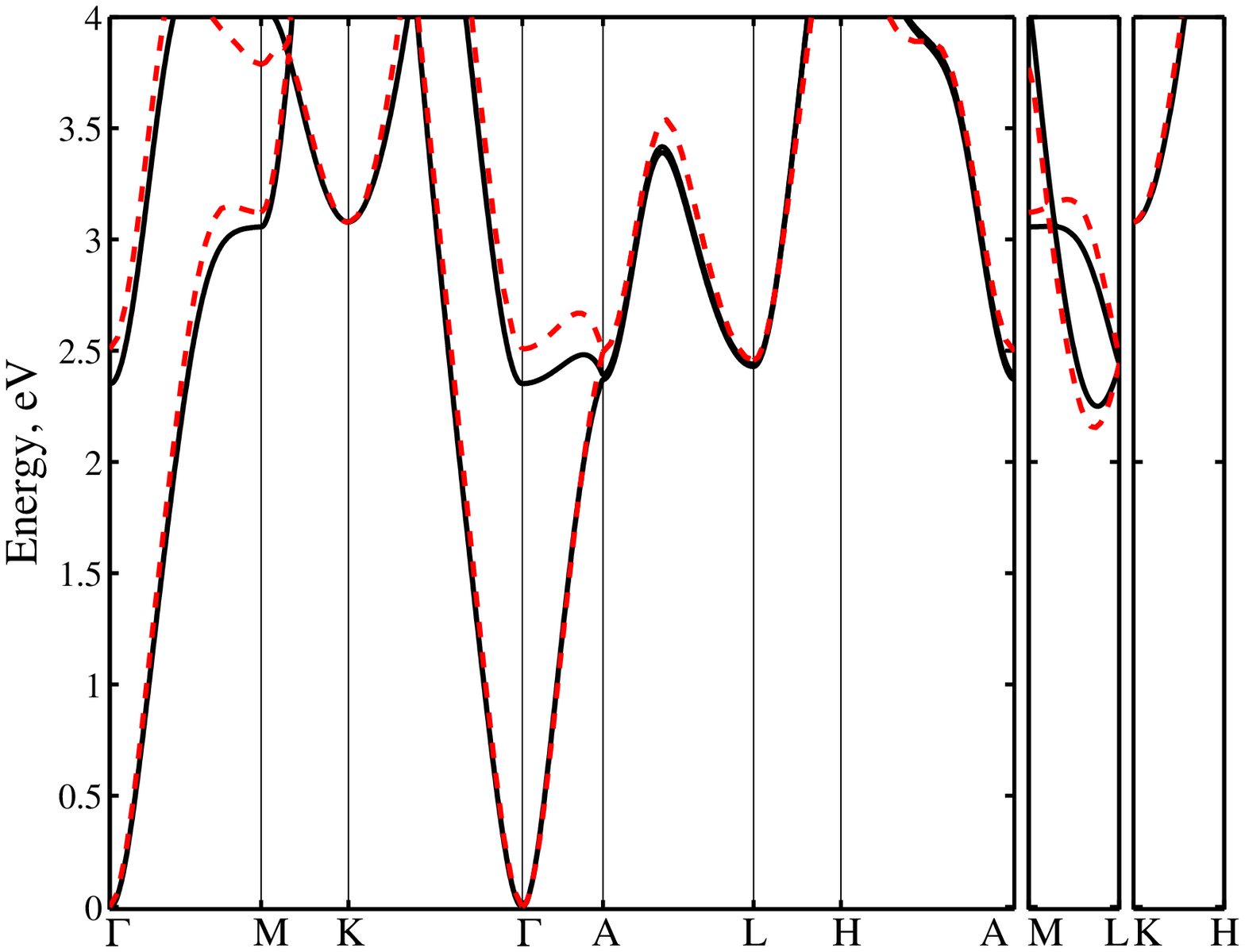}}
\centerline{(a)}
\centerline{\includegraphics*[width = 0.70\linewidth]{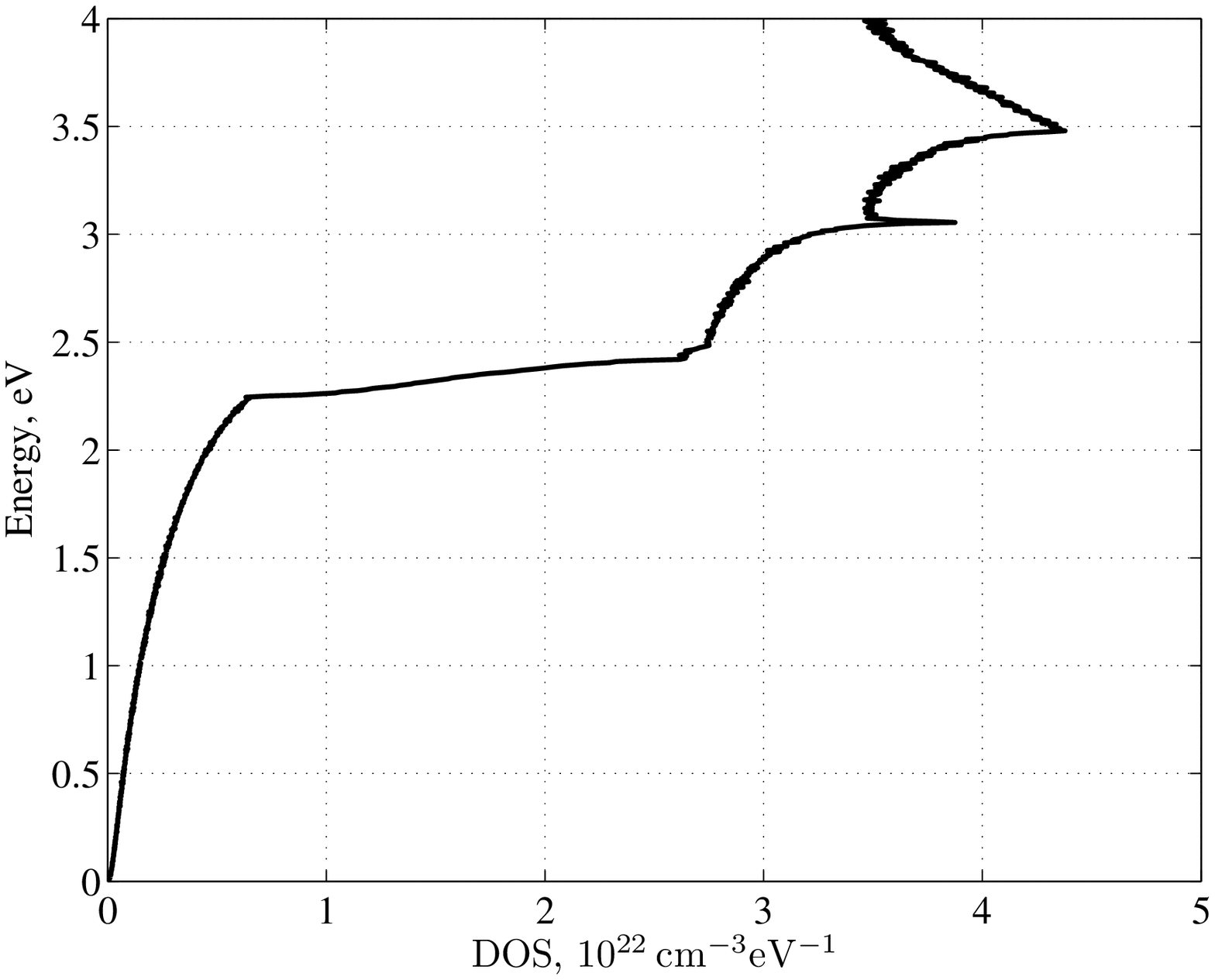}}
\centerline{(b)}
\caption{(a) CB of GaN computed with NL-EPM (black solid lines) and DFT-HSE (red dashed lines).
(b) Corresponding density of states determined from the NL-EPM electronic structure.}
\label{f:electronic_structure}
\end{figure*}

Fig.\,\ref{f:electronic_structure}(a) shows the details of the CB in GaN, as computed with the nonlocal empirical pseudopotential method (NL-EPM) \cite{2000Goano_JAP.1, 2007Bellotti_JAP} and with density functional theory based on hybrid functionals (DFT-HSE) \cite{2003Heyd_JCHP, 2003Heyd_JCHP.erratum} as implemented in the \textsc{VASP} code \cite{1996Kresse_PRB, 1999Kresse_PRB}.
(The Hartree-Fock contribution to the hybrid functional was 28\%.)
These results are in good agreement with each other and with state-of-the-art \emph{ab initio} calculations (see e.g.\ \cite{2009Delaney_APL, 2010Svane_PRB, 2011deCarvalho_PRB}).
According to the NL-EPM bands, which will be used in the following transport analysis, the lowest secondary valley in the CB lies along the L--M segment at 2.25\,eV above $\Gamma_1^c$, while L$_1^c$ is a saddle point, and higher local minima are at 2.35\,eV ($\Gamma_3^c$) and 3.1\,eV (K$_2^c$) above the CB minimum.
The NL-EPM density of states (DOS) of the CB is reported in Fig.\,\ref{f:electronic_structure}(b), where the peaks corresponding to the satellite valleys are clearly visible.

\section{Hot electron transport in GaN}

\begin{figure}
\centerline{\includegraphics*[width = 1.00\linewidth]{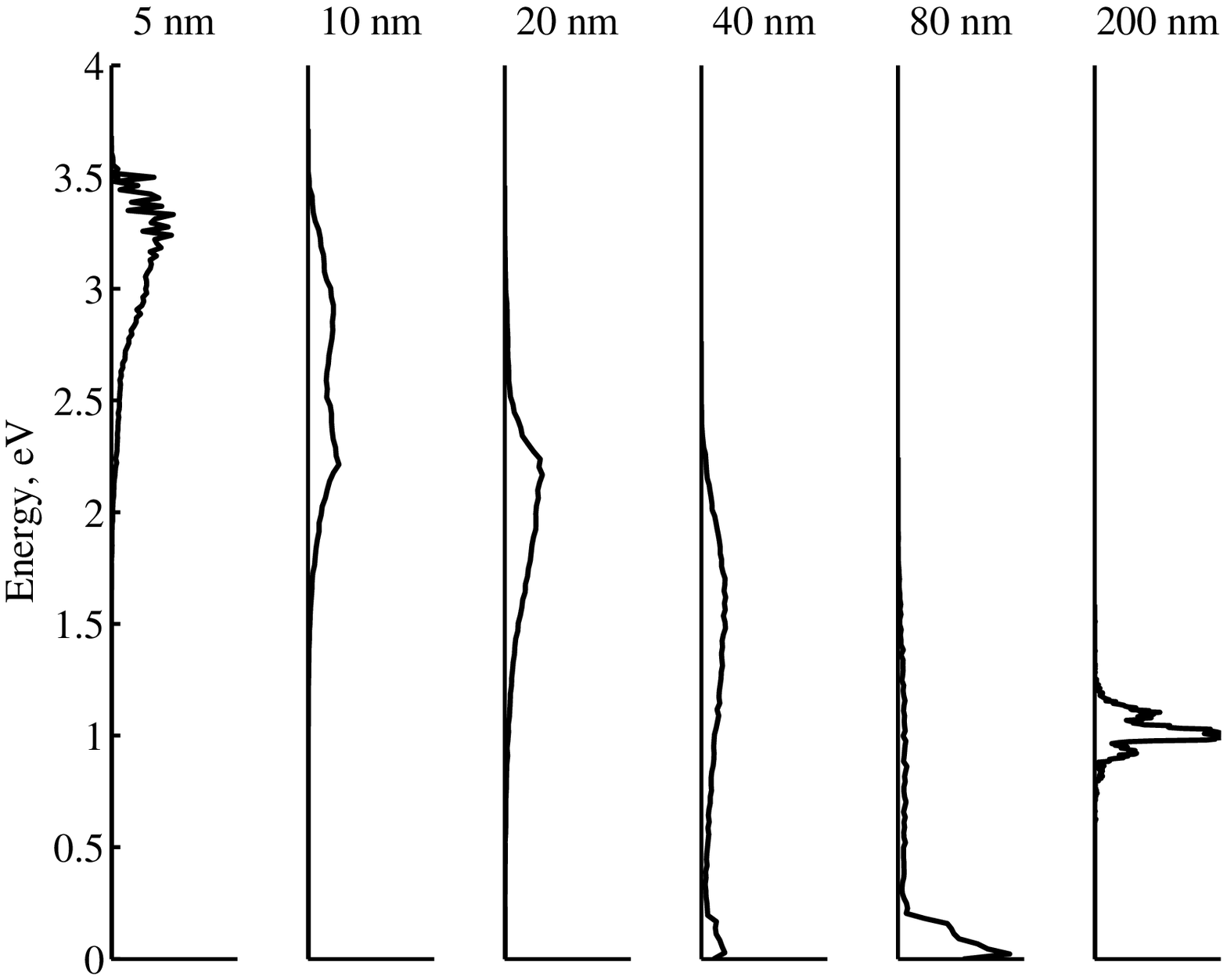}}
\caption{Kinetic energy distribution of electrons in the GaN $p$-cap, as simulated with FBMC, at increasing distance from the interface with the EBL.}
\label{f:EDC}
\end{figure}

\begin{figure*}
\centerline{\includegraphics*[width = 0.35\linewidth]{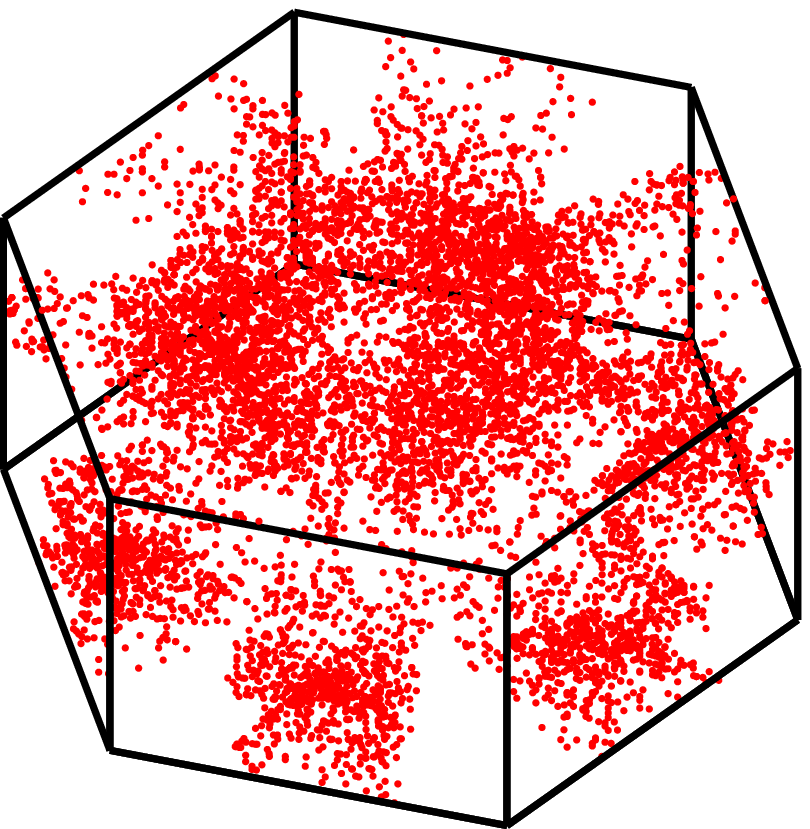}}
\centerline{\includegraphics*[width = 0.35\linewidth]{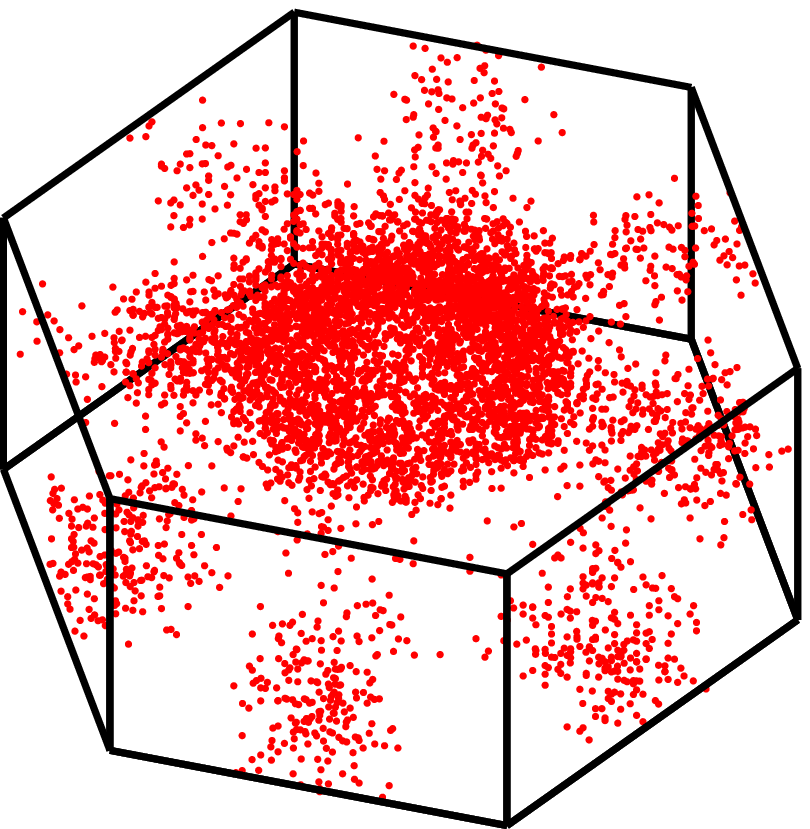}}
\centerline{\includegraphics*[width = 0.35\linewidth]{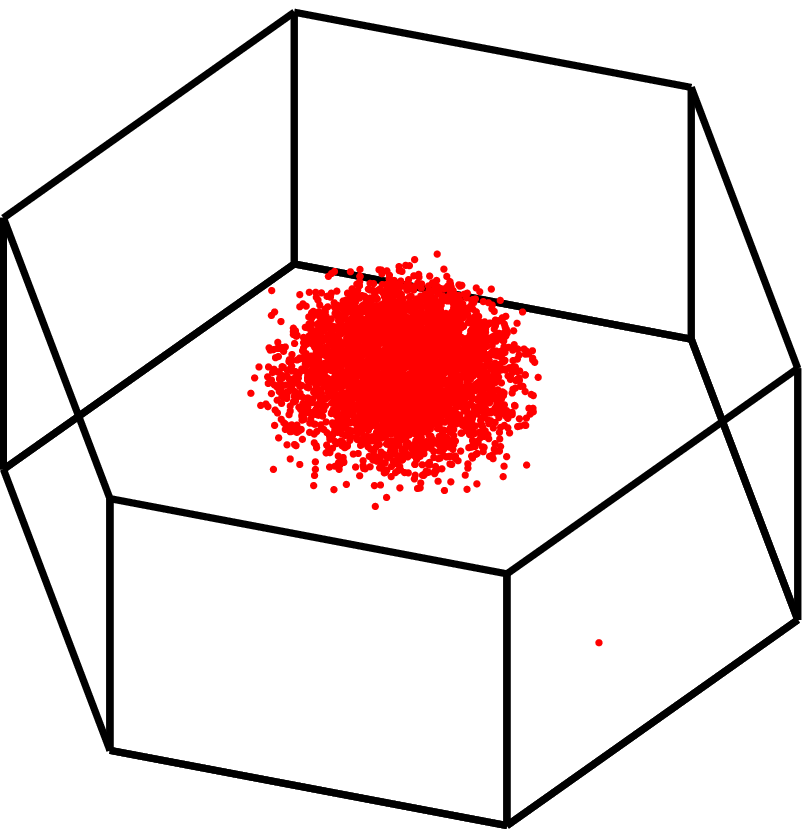}}
\caption{Electron distribution in the Brillouin zone, as simulated with FBMC, in the GaN $p$-cap at 5\,nm, 10\,nm and 40\,nm from the interface with the EBL.
In the first two cases, most carriers still populate the satellite valleys located in K$_2^c$, $\Gamma_3^c$, and along the L--M segment, whereas at a distance comparable with the EBL thickness the distribution is already dominated by carriers in the $\Gamma_1^c$ valley at the bottom of the CB.}
\label{f:distrib_k}
\end{figure*}

We investigated hot electron transport in the GaN $p$-cap of the LED studied in \cite{2013Iveland_PRL} using a full-band Monte Carlo (FBMC)  simulator \cite{2009Bertazzi_JAP} that incorporates the full details of the NL-EPM bands reported above.
Unlike conventional ``many-valley'' Monte Carlo descriptions of carrier tranport \cite{2001Farahmand_TED, 2012Guetle_SST, 2013Hadi_JAP}, whose results are determined by several poorly known empirical parameters (effective mass tensors and nonparabolicity coefficients for each satellite valley, intervally phonon energies and deformation potentials), FBMC does not rely on local approximations of the band structure and of the lattice dynamics, and provides a unified treatment of both intra- and inter-valley scattering processes.
In particular, nonpolar carrier-phonon interactions are computed within the framework of the rigid pseudoion approximation, where the phonon dispersion relation is determined with \emph{ab initio} techniques.
Band-to-band tunneling effects are also included in the framework of the formalism proposed by Krieger and Iafrate \cite{1986Krieger_PRB, 2001Nilsson_JAP}
\footnote{FBMC accurately predicts the multiplication gain and noise properties of GaN-based avalanche photodiodes (APDs) \cite{2009Moresco_JAP} and nonstationary transport effects observed by time-resolved electroabsorption measurements \cite{2003Wraback_APL}.}.

Transport through the GaN $p$-cap of hot electrons due to Auger recombination processes in the active region was studied by simulating the injection, immediately above the heterointerface between EBL and $p$-cap, of electrons with energies higher than the satellite valleys of the GaN CB (L--M, $\Gamma_3^c$, K$_2^c$) and with wave vectors randomly distributed in the whole Brillouin zone, as expected if Auger recombination is phonon-assisted.

Fig.\,\ref{f:EDC} reports the simulated EDCs in the GaN $p$-cap at increasing distance from the interface with the EBL.
The corresponding electron distributions in the Brillouin zone are shown in Fig.\,\ref{f:distrib_k}.
It may be observed that hot electrons thermalize in the $\Gamma_1^c$ valley at the bottom of the CB (i.e. at energies much smaller than the observed higher-energy peak) in less than half the $p$-cap thickness before reaching the BBR.

The FBMC results presented in Fig.\,\ref{f:EDC} are probably conservative, because they neglect (1) carrier thermalization in the $40\,$nm thick EBL and (2) carrier-carrier scattering mechanisms, which would further accelerate the relaxation process \cite{1991Mosko_PRB}.

These results suggest that the higher-energy peaks in the measured EDCs are probably uncorrelated with the carrier distribution in the active region.
This would not imply that Auger recombination, and possibily Auger-induced leakage \cite{2012Deppner_PSSR}, play a negligible role in LED droop, but that an Auger signature cannot be recovered from the experiment presented in \cite{2013Iveland_PRL} if our understanding of the LED structure under investigation is correct.

\section{Carrier heating by the electric field in the BBR}

\begin{figure*}
\centerline{\includegraphics*[width = 0.70\linewidth]{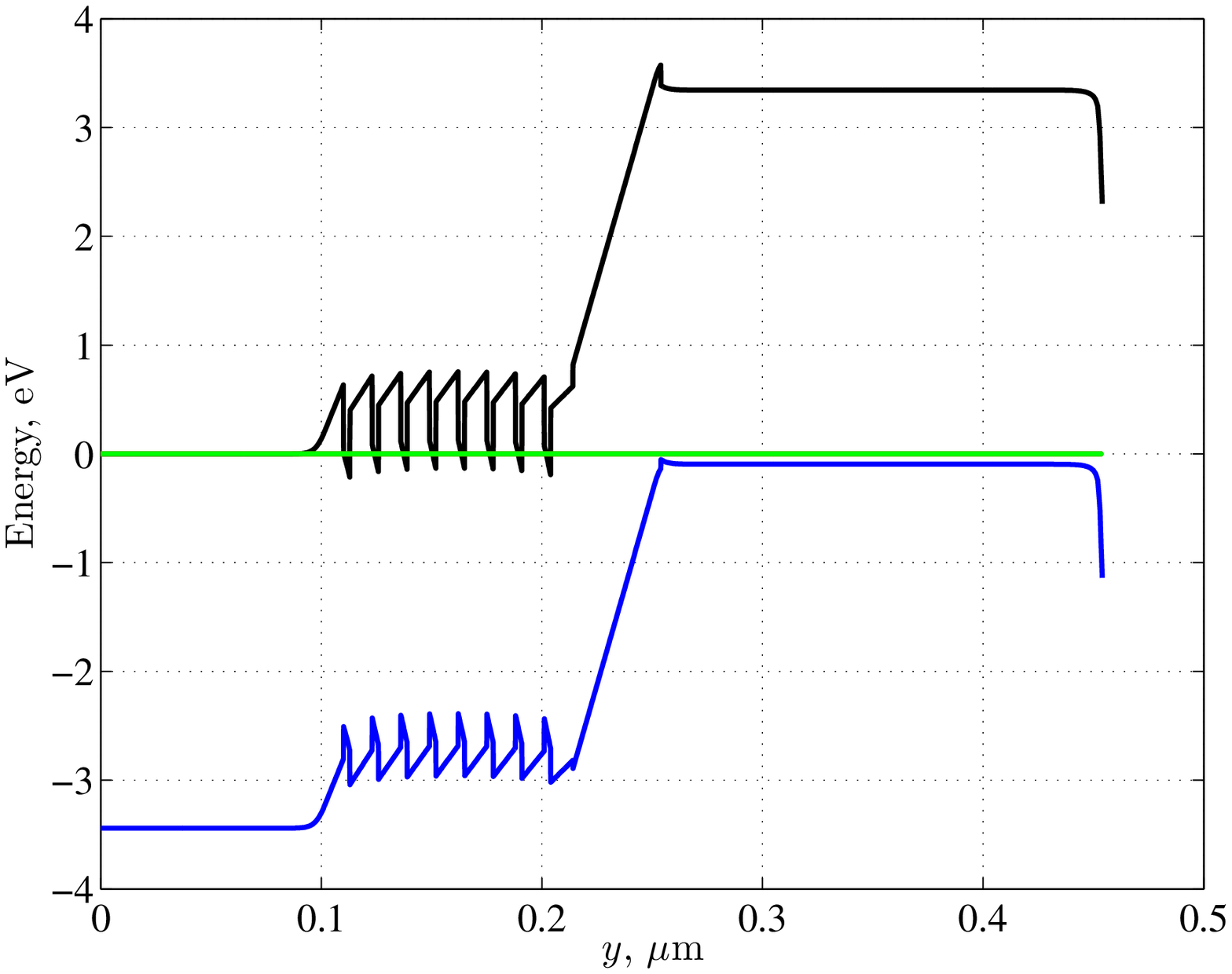}}
\centerline{(a)}
\centerline{\includegraphics*[width = 0.70\linewidth]{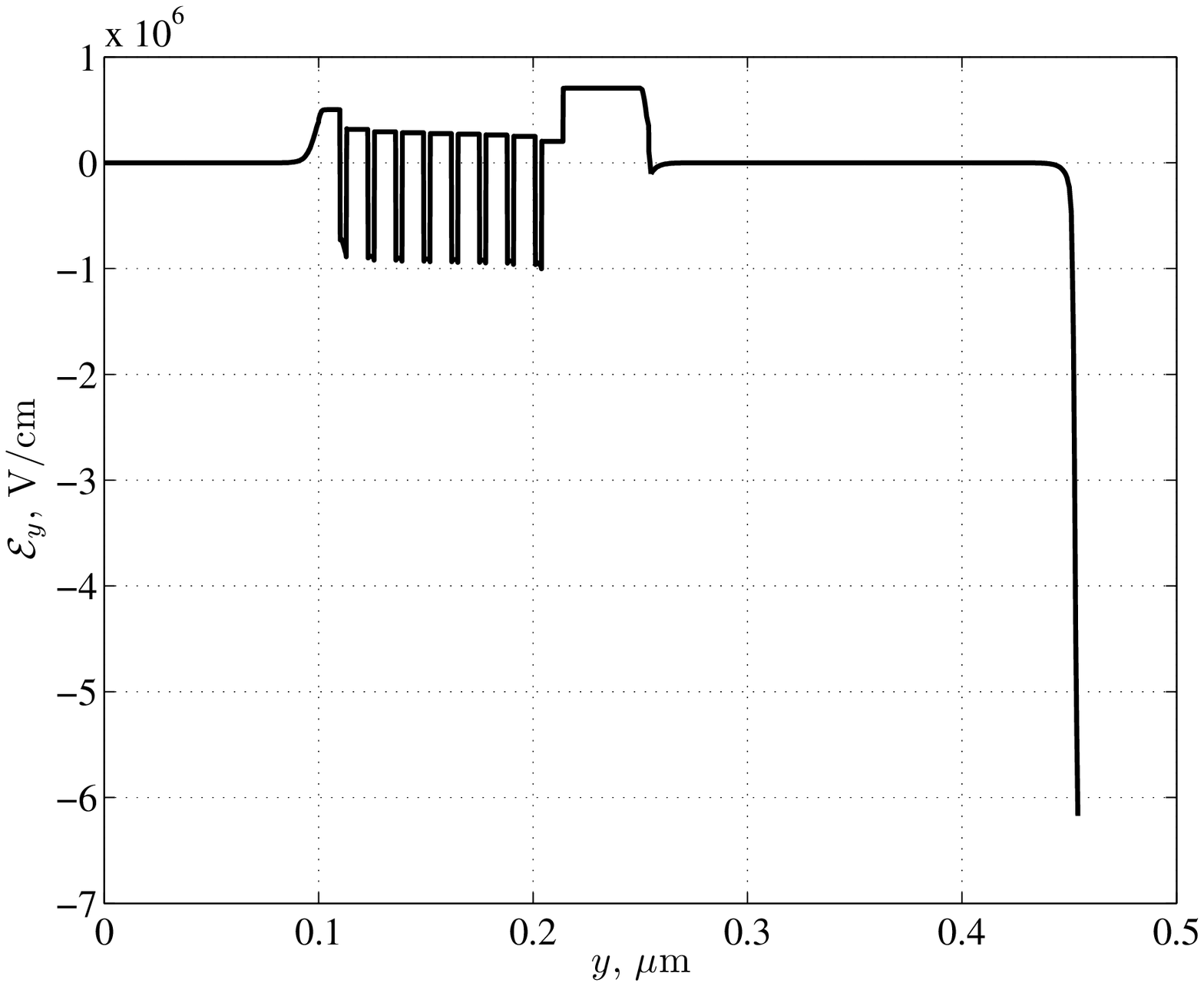}}
\centerline{(b)}
\caption{(a) Band diagram and (b) vertical component of the electric field across the LED structure studied in \cite{2013Iveland_PRL} at equilibrium, assuming a conservative band bending of 1\,eV at the cesiated GaN surface in accordance with the estimated $\Phi_\text{GaN} = 2.3$\,eV.}
\label{f:bands_Efield}
\end{figure*}

The physical mechanisms responsible for the observed bimodal EDC should be identified in order to assess whether the technique proposed in \cite{2013Iveland_PRL} can be actually used to probe the energy distribution function of carriers in the active region of an appropriately designed GaN-based LED.
An explanation of the experiment would require more details about the complex device under study, in particular at the $p$-cap surface, in order to allow a realistic two- or three-dimensional analysis, and about the Mg doping profile to assess possible surface segregation \cite{2002Figge_APL}.
From the information provided in \cite{2013Iveland_PRL}, we can speculate that carrier heating by the electric field in the BBR could account for the higher-energy peak.

The electric field profile across the GaN $p$-cap used in the FBMC analysis was computed with a drift-diffusion simulator (\textsc{APSYS} from Crosslight Software Inc.) taking into account incomplete dopant ionization \cite{2011Chiaria_JQE} and assuming a conservative band bending of about 1\,eV at the cesiated surface \cite{1999Wu_JAP, 2004Liu_APL} in accordance with the work function $\Phi_\text{GaN} = 2.3$\,eV determined in \cite{2013Iveland_PRL}.
With an Mg density in the $p$-cap $N_A \approx 1.8 \cdot 10^{20}$\,cm$^{-3}$, as estimated in \cite{2013Iveland_PRL}, and an acceptor ionization energy $E_A \approx 200$\,meV, the calculated BBR has an extension of about 10\,nm (comparable with the mean free path in GaN) and the vertical component of the electric field has a peak value of about 6\,MV/cm at the $p$-cap surface (see Fig.\,\ref{f:bands_Efield}).
Both values are largely independent from the applied bias, since they are essentially determined by the complete ionization of the Mg impurities near the $p$-cap surface because of the cesiation-induced band bending.

Thermalized electrons approaching the BBR are accelerated by the large electric field and propagate quasi-ballistically to the surface.
The resulting simulated EDC at the $p$-cap surface (Fig.\,\ref{f:EDC}, 200\,nm) exhibits a single peak lying at a kinetic energy corresponding to the total band bending.
The energy of this peak is unrelated with the energy of the satellite valleys in the CB (according to the FBMC simulation, all electrons in the BBR are still in the $\Gamma$ valley).
(Very similar EDCs, not reported here, were observed also when considering in the simulations, with the same band bending at the cesiated surface, lower Mg densities in the $p$-cap down to $N_A \approx 2 \cdot 10^{18}$, corresponding to electric field peaks always in the MV/cm range.)

If this description is correct
\footnote{As an additional speculation, the higher-energy peak in the EDC was not observed in \cite{2013Iveland_PRL} under external optical excitation at energy close to the LED emission possibly because  electron photoemission does not occur, as suggested by Iveland \emph{et al.}, in the GaN BBR (which is almost transparent to the electroluminescence in the active region and is depleted of both electrons and holes if the LED is not above threshold) but only in the Cs activation layer, where it would originate the measured lower-energy peak.},
the higher-energy peak observed in the experimental EDCs could be ascribed to electrons reaching the BBR after escaping from the active region either because of Auger-induced leakage, or through other leakage/flyover mechanisms \cite{2012Saguatti_TED}; e.g.,
\textsc{APSYS} simulations suggest that thermionic leakage could be significant in the LED under test even in the presence of relevant Auger recombination.

\begin{acknowledgments}
The authors would like to thank Prof.~Bernd Witzigmann, Dr.~Matteo Meneghini, Michel Lestrade and Marco Vallone for useful discussions.
\end{acknowledgments}

\end{document}